# Fair Efficiency Comparisons of Decoy-state Quantum Key Distribution Protocols

LI Hong-xin[1,2], GAO Ming[2], YAN Xue-ping[1], YAN Bao[1], HAN Yu[1], SHAN Ling[3],MA Zhi[2]

(1. Department of Language Engineering, Luoyang University of Foreign Languages, Luoyang, Henan, 471003, China)

(2. State Key Laboratory of Mathematical Engineering and Advanced Computing, Zhengzhou, 450001, China）

(3. College of Animal Science and Technology, Henan University of Science and Technology, Luoyang, Henan, 471003, China）

**Abstract**： Secure key rate of decoy-state quantum key distribution protocols has been improved with biased basis choice, however, the security standards and parameters of current protocols are different. As a result, we cannot give an accurate key rate comparison between different kinds of protocols. Taking the schemes based on different formula of secure key rate as examples, we give a fair comparison between typical protocols under universal composable security standard. Through analyzing the relationship of security parameters in post-processing stage and final secure key, we achieve the unified quantification between protocols based on Gottesman-Lo-Lütkenhaus-Preskill formula and the ones under universal composable security. Based on the above research, the impact of different sending length and secure parameters on secure key rate is investigated, meanwhile, we give the dependent relationship between secure key rate and sending length under different secure parameters. Besides, we analyze the importance and conditions of fair comparison. For the first time we give a fair comparison between the protocols based on GLLP formula and smooth entropy, and taking Raymond protocol and Toshiba protocol as examples, we analyze the way for improving secure key rate in the light intensity choice and the single bit error rate estimation method.

## 1 Introduction

Quantum key distribution (QKD) provides unconditional secure shared key for communication parties based on quantum mechanics. Under ideal conditions, the ways of security proof mainly include uncertainty principle[1-3], entanglement distillation protocol (EDP) [4-6] and information theory[7]. However, the ideal conditions are difficult to achieve in practical QKD systems. For the imperfect cases of practical QKD systems, Gottesman et al. conducted a deep research on security analysis[8] and presented GLLP formula based on the EDP protocol, the security of which is equal to BB84 protocol[9]. In order to securely apply the final key in message encryption, Mayers[10] and Renner[11] brought universal composable (UC) frame from classical cryptography into QKD schemes based on BB84 protocol, describing indistinguishability between the ideal and practical secure schemes by trace distance. Besides, for the potential security danger brought by multi-photon pulses and practical quantum channel loss, Hwang[14], Wang[15] and Lo[16], respectively presented decoy-state idea and practical decoy-state QKD schemes, greatly improving the secure key rate. Communication parties transfer finite data since the communication time is limited in realistic conditions. As a result, the related research of decoy-state QKD postprocessing schemes is gradually turning into finite-key schemes [17-33].

Current decoy-state QKD protocols, provide many different ways based on light intensity choice, key generating part, bias basis setting and so on, claiming that they have improved secure key generation rate to some extent.

In ref. [22], they use Z basis of signal state for generating the final key, choose the signal state's light intensity smaller than decoy-state, and set the probability of choosing X basis for decoy state as 1. They can generate the final secure key under the length of $10^6$ bits of receiving code.

Ref. [24] chooses the signal state's light intensity larger than decoy-state, and sets the probability of choosing Z basis for signal state as 1. Compared with standard BB84 protocol, its secure key rate is improved by 45 percent.

In ref. [32], they generate key in both X basis and Z basis of signal state, and claim that the scheme had so far achieved the highest key rate compared with other experimental ones.

However, the researches concerning the improvement of secure key rate mostly pay attention to the simple comparison between the protocol and standard BB84 protocol on the infinite-key condition [15, 16, 17, 24, 33]. There are few papers about efficiency comparisons between protocols based on different security standards and parameters.

In ref. [24], the formula of secure key rate based on GLLP formula, just including the security parameters of some postprocessing stage and considering other stages as perfect, is not overall.

Ref. [22, 23, 34] claim that their protocols are under UC security standard all, however, when it comes to the security parameters of postprocessing stages, their analysis is different and so is the UC security standard definition. For example, in ref. [22], three parameters of failure probability and parameters of smooth entropy in the postprocessing stage are defined as its UC security standard parameters. Ref. [33] interprets it as the sum of security and correctness parameters of final key, and security parameters description is based on trace distance. As for Ref. [24], they reaesrch parameters of failure probability in postprocessing stages by quantum fidelity.

Therefore, a standard and platform of fair comparison is needed urgently to test secure key rate of decoy-state QKD protocols so that the improvement of secure key rate of decoy-state QKD protocols with biased basis choice is provided reference.

Considering the overall factors influencing the secure key rate is the difficulty of fair efficiency comparison. As a result of it, for the new decoy-state QKD protocols of biased basis choice raised in the last two years, and according to different features in formula of secure key rate, light intensity choice, bias basis setting and more, our article uses typical protocols to analyze. Firstly we research the unified quantification relationship of security parameters between decoy-state QKD schemes based on GLLP formula and protocols under UC security standard, and achieve the conversion of protocols from non-UC security to UC security. On this basis, we research the influence of different sending length and security parameters on the secure key rate using the protocol based on GLLP formula under UC security standard (UC-Wei protocol). Meanwhile, taking the protocol based on smooth entropy (UC-Raymond protocol) as examples, our article analyzes the dependent relationship between secure key rate and sending length under different security parameters. Then on the condition of the same finite-key length，security parameters and statistical fluctuation method, we analyze the importance and conditions of fair comparison. We note that the comparison in this article is relatively fair.

For the first time we give a fair efficiency comparison between decoy-state QKD protocols with biased basis choice based on GLLP formula and protocols based on Devetak-Winter bound[35] and smooth entropy, and this is the main contribution of this article. After that we analyze the advantage brought by involving decoy state in key generation on the basis of UC-Raymond protocol, which uses single signal sate to generate key. Finally we take Toshiba protocol（T12 protocol）[32] as an example to research advantages and disadvantages between different single bit error rate estimation methods. Because the comparisons in our article take contrast effects between different key lengths into consideration, conclusions can provide reference to the application of decoy-state QKD protocols.

The article is organized as follows. Section II introduces the definition of UC security standard and UC security parameters, and analyzes the unified quantification interrelation between protocols under non-UC security and the ones under UC security standard. Section III researches the relationship of sending length, security parameters and secure key rate. Section V discusses the importance and conditions of fair efficiency comparison, and its point is in three aspects. We conclude in VI.

## 2 Analysis of security standard

2.1 UC security standard

UC security definition [8]: in the decoy-state QKD protocols under the finite-key condition, and assuming the adversary, Eve, adopts the optimal attack strategy, the final key is called $\varsigma$–indistinguishable with unconditionally secure key, or, $\varsigma$–UC secure, if $\forall \varsigma \geq 0$,

$$\min_{\rho_E} \frac{1}{2} \| \rho_{SE} - \tau_S \otimes \rho_E \| \leq \varsigma, \qquad (1)$$

where the trace distance $\frac{1}{2}\|\rho-\sigma\| = \frac{1}{2}Tr|\rho-\sigma|$ is the maximum probability of distinguishing the two quantum states ρ and σ, and $\rho_{SE}$ is the classic & quantum mixed state of Alice's classical key S and the eavesdropper, Eve's quantum state, E. $\tau_S$ denotes the fully mixed state on S and $\rho_E$ is the Eve's quantum state.

2.2 UC security parameter

Security parameter is generally considered as the failure probability of the whole QKD protocol or some postprocessing stage. Regarding $\varsigma$ in the UC security definition as the UC security parameter of final key of decoy-state QKD protocols, and assuming that $\varepsilon$ is the failure probability of protocols under UC security standard in the post-processing stage, we can get $\varsigma \geq \varepsilon$.

QKD postprocessing procedure [34] contains five elements mainly, including basis sifting, bit error correction, error verification, parameter estimation and privacy amplification. Among them the basis sifting stage is mostly finished in an authenticated classical channel and is thought as ideal. We assume that the others have failure probability under UC security standard, and in a sense the smooth entropy parameter is a failure probability [8]. In the current decoy-state QKD protocols under UC security standard, the formula of secure key rate are based on uncertainty reletion[33] between Devetak-Winter bound[22,32] and smooth entropy, so its analysis of failure probability in the postprocessing procedure and smooth entropy parameters are based on trace distance directly. While in the decoy-state QKD protocols based on GLLP formula [24], we just focus on the failure probability of random sampling in phase error rate estimation step under the finite-key condition, under non-UC security. Because GLLP formula derivation is on the basis of EDP protocol, its security parameter description is based on quantum fidelity

2.3 Security parameter quantization

We assume that $\varepsilon_{UCBC} \geq 0$, $\varepsilon_{UCEV} \geq 0$, $\varepsilon_{UCPE} \geq 0$, $\varepsilon_{UCPA} \geq 0$ and $\bar{\varepsilon}_{UC} \geq 0$ are respectively failure probability of bit error correction, error verification, parameter estimation, privacy amplification and smooth entropy parameter in the postprocessing procedure in the decoy-state QKD protocols with smooth entropy on the finite-key condition under UC security standard, and that they satisfy $\varepsilon_{UCPE} < \bar{\varepsilon}_{UC}$. Besides, define $n_{SEPE}$ is numbers of parameters needing estimation. For that the description of related secure parameters is based on trace distance, the UC security parameter of final key can be defined as the sum of the whole security parameters in the postprocessing stage and the smooth entropy parameter: $\varsigma_{UC} = \varepsilon_{UC} + \bar{\varepsilon}_{UC} = \varepsilon_{UCBC} + \varepsilon_{UCEV} + n_{PE}\varepsilon_{UCPE} + \varepsilon_{UCPA} + \bar{\varepsilon}_{UC}$.

Taking Wei protocol as an example, apart from the failure probability of phrase error estimation in parameter estimation, the one in the postprocessing stage is also supposed to be taken into consideration, to achieve the quantification from decoy-state QKD protocol based on GLLP formula to the protocols under UC security standard. Similarly, we define $\varepsilon_{GBC} \geq 0$, $\varepsilon_{GEV} \geq 0$, $\varepsilon_{GPE} \geq 0$ and $\varepsilon_{GPA} \geq 0$ are respectively failure probability of the last four elements above, and $n_{GPE}$ is the number of parameters needing estimation except phase error rate, and in Wei protocol it is 4, thus the whole security parameter in the postprocessing stage is $\varepsilon_{GLLP} = \varepsilon_{GBC} + \varepsilon_{GEV} + n_{GPE}\varepsilon_{GPE} + \varepsilon_{GPA} + \varepsilon_{Gph}$.

According to the quantification relationship between quantum fidelity and trace distance[34], $Tr|\rho-\theta| \leq \sqrt{1-F(\rho,\theta)^2}$, where $F(\rho,\theta) = Tr\sqrt{\rho^{\frac{1}{2}}\sigma\rho^{\frac{1}{2}}}$ is the quantum fidelity between quantum states $\rho$ and $\theta$, we can get the correspondence equation $\varsigma_{GLLP} = \sqrt{\varepsilon_{GLLP}(2-\varepsilon_{GLLP})}$ between $\varepsilon_{GLLP}$, the failure probability in the postprocessing stage in Wei protocol, and $\varsigma_{GLLP}$, the UC security parameter of final key, and it is an important basis for the analysis of of security parameters between two protocols under UC security standard. Therefore, on condition that the UC security parameters of final key are the same in two protocols, the quantitative equation is

$$\varepsilon_{UC} + \bar{\varepsilon}_{UC} = \varsigma_{UC} = \varsigma_{GLLP} = \sqrt{\varepsilon_{GLLP}(2-\varepsilon_{GLLP})} \qquad (2)$$

## 2.4 Security parameter value choice

The security of a QKD protocol can be showed by UC security parameter values of final key, and the smaller parameter represents the more secure protocol. For example, the parameter of final key in the Raymond protocol[22] is $10^{-5}$, and comparing it with the one in the T12 protocol, $10^{-10}$, which has been realized in experiment, we find it obvious that the latter security is better. Besides, secure key rate of finite-key protocol is influenced by security parameter choice to an extent. Taking UC-Wei protocol and UC-Raymond protocol as examples, we research the influence on the secure key rate between different key length and security parameter choice in detail in the third part.

## 3 Security parameter and finite-key length

Firstly taking Wei protocol in Ref. [24] as an example, the influence on secure key rate of protocols based on GLLP formula under UC security standard of security parameter value choices is analyzed. The formula of UC-Wei protocol reads:

$$R_{Wei} \geq \frac{N_\mu p_{Bz}}{N}\left\{-fQ_\mu H(E_\mu)+Q_1^z[1-H(e_1^{pz})]+Q_0\right\}-\frac{k_{WEV}+k_{WPA}}{N} \quad (3)$$

where because error verification stage uses failure probability $\varepsilon_{GEV}$ to guarantee the high consistance of bit of communication parties and that if this verification fails, the parties either go back to the bit error correction step or abort the QKD process, so we define the error correction stage ideal. Besides, error verification stage, which is performed by using sets of Toeplitz matrices, consumes key bits $k_{WEV}=1+\log_2\frac{NP_\mu q_{Bz}}{\varepsilon_{GEV}}$. Similarly $k_{WPA}$ is secret key bits consumed in privacy amplification stage, and $k_{WPA} \approx 1+\log_2\frac{NP_\mu q_{Bz}+l-1}{\varepsilon_{GPA}}$, where $l$ is the length of final key, and $l=N_\mu p_{Bz}\left\{Q_1^z[1-H(e_1^{pz})]+Q_0\right\}-\log\frac{1}{2^{-(NP_\mu p_{Bz}-l)}}$. Define $\varepsilon_{Gph}$ is the failure probability of phase error estimation and satisfies $\varepsilon_{Gph} \approx P_{W\theta x} \geq 0$, where $P_{W\theta x} \equiv \Pr\{e_{pz} \geq e_{bx}+\theta_x\}$ is the failure probability of phase error estimation in the Z basis, $e_{pz}$ is the single phase error rate in the Z basis, $e_{bx}$ is the single bit error rate in the X basis, and $\theta_x$ represents the deviation between $e_{pz}$ and $e_{bx}$. Intuitively, while security parameters in error verification and privacy amplification are smaller, the bits consumed are larger. In the following, to research the statistical fluctuation effect of measuring parameters under finite key condition conveniently, we will focus on the failure probability of protocols using statistical fluctuation.

To check the effect of finite-key length and security parameter choice on secure key rate, we define the finite-key length $N \in \{6\times10^9, 10^{12}, 10^{15}\}$, where the finite-key length refers to sending quantum bit length, and $6\times10^9$ is the shortest sending length to generate key in UC-Wei protocol. Then we choose the relatively secure UC security parameter of final key, $\varsigma_{Wei} = 1\times10^{-10}$, and the one of original Wei protocol, $P'_{W\theta x} \approx \varepsilon'_{GPE} = 5.73\times10^{-7}$, which corresponding $\varsigma'_{Wei} \approx 2.5\times10^{-3}$. The statistical fluctuation method is still standard error analysis, and other practical system parameters remain constant. Figure 1(we use figure 1.1, 1.2, 1.3 and 1.4 to denote the above left, the above right, the lower left and the lower right graph) is the contrast diagram of secure key rate between different finite-key length when the security parameter is the same or different security parameter while the length is the same, where $t$ represents channel transmission loss, $R$ is secure key rate, and $sp$ is UC security parameter value choice.

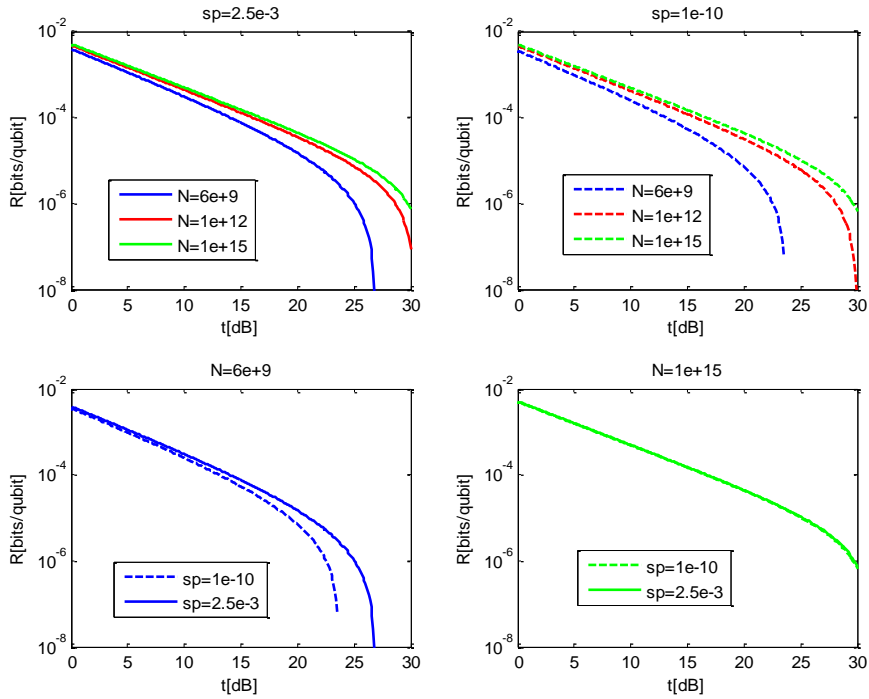

Figure 1: contrast diagram of security key rate

Figure 1.1 and 1.2 show changes of related secure key rate of two security parameters.

Through numerical optimization, we can find that when the finite-key length $N = 1\times10^{15}$, the curves have a steady decline trend in two subgraphs along with the increase of channel loss. While the finite key is shorter, the influence of statistical fluctuation becomes greater, and so the decline trend is quicker.

Figure 1.3 and 1.4 are contrast diagrams of related secure key rate between different UC security parameters when $N=6\times10^9$ and $N=1\times10^{15}$.

We can conclude from figure 1.3 that when UC security parameter $\varsigma_{Wei}=1\times10^{-10}$, secure key rate is lower. It is because $\varsigma_{Wei}=1\times10^{-10}$ is corresponded with the major security parameter in the post-processing stage, $P_{W\theta x}\approx\varepsilon_{GPE}=1\times10^{-21}$, which is less than $P'_{W\theta x}\approx\varepsilon'_{GPE}=5.73\times10^{-7}$ far in the original protocol, and it results that in the phrase error rate estimation and standard error analysis，deviation's and standard error's multiples are larger，with $\theta_{Wx}>\theta_{Gx}$ and $u_{W\alpha}=9.5>u_{G\alpha}=5$，where $\theta_{Wx}$ and $\theta_{Gx}$ is respectively the deviations of phrase error rate estimation in two parameter values, and $u_{W\alpha}$ and $u_{G\alpha}$ are respectively the standard error multiples when using standard error analysis. We can see that the security parameter value choice can influence secure key rate in a degree, and when the security command is higher, the related rate is lower.

In the figure 1.4 we can see that the curves in different security parameters almost coincide. This is to say, when the sending length is comparatively long, the influence on secure key rate of security parameter can be ignored. Taking $N=1\times10^{15}$ as an example, we choose the channel loss $t=25dB$, and the ratios between deviation and measuring values in statistical fluctuation, and the ratios between the error of secure key rate resulted from the deviation and key rate with infinite-key length are shown in table 1.

Table 1 ratios of related parameters

| ratio \ security parameter | $\Delta Q_\mu$ | $\Delta Q_\nu$ | $\Delta Y_0$ | $\Delta Q_0$ | $\Delta\theta_x$ | $\Delta R$ |
|---|---|---|---|---|---|---|
| $2.5\times10^{-3}$ | $3.6\times10^{-2}$ | $2.0\times10^{-5}$ | $1.2\times10^{-3}$ | $2.0\times10^{-3}$ | $2.5\times10^{-3}$ | 1.8% |
| $1\times10^{-10}$ | $4.8\times10^{-2}$ | $3.7\times10^{-5}$ | $1.8\times10^{-3}$ | $3.4\times10^{-3}$ | $4.3\times10^{-3}$ | 2.5% |

In table 1 $\Delta Q_\mu$, $\Delta Q_\nu$, $\Delta Y_0$, $\Delta Q_0$, $\Delta\theta_x$ and $\Delta R$ respectively represent the ratios between the deviation and measuring values, and the proportion that the difference resulted from these deviations accounts for in the secure key rate on the infinite-key length in phrase error rate, signal state detection rate, decoy-state detection rate, vacuum state detection rate and the detection rate of vacuum pulse in signal state. From these ratios, we can conclude that the proportion of deviation of relevant parameters occupying in the measuring values is rather small, and that the proportions of difference of secure key rate resulted 1.8% and 2.5%, are less than the ones (26.8% and 35.0%) when $N=6\times10^9$ far. The analysis above shows that when the sending length is comparatively long, the effect of secure parameter value choice and related statistical fluctuation can decrease to a pretty small order of magnitude. Therefore, in practical QKD systems, we can increase sending length properly to reduce the influence of statistical fluctuation. For example, assume sending length $N = 1 \times 10^{15}$, and $\varsigma_{Wei} = 1 \times 10^{-10}$, $\varsigma'_{Wei} \approx 2.5 \times 10^{-3}$, both of ratios of secure key rate, $9.9823$ and $9.9750$, is close to 1.

The analysis above shows the dependent relationship between secure key rate and sending length when security parameters are different. In other word, a right length exits which can reduce the influence of statistical fluctuation greatly. To confirm further the reasonableness of the conclusion, we uses the protocol based on smooth entropy (Raymond protocol) to this relationship.

The secure key rate formula in Raymond protocol is:

$$R_{Raymond} = P_\mu Q(\mu) q_z^2 [S_{\xi_\mu}(A|E,\mu) - \Delta(n_\mu) - \frac{leak_{EC}(e_X(\mu))}{n_\mu}] \qquad (4)$$

where see parameter definition and statistical fluctuation method in Ref.[22]. we emphasize that because the bit error rate estimation uses failure probability $\varepsilon_{UCBC} \geq 0$ to guarantee the high consistance of bits of communication parties, we suppose error verification stage is ideal.

Figure 2 shows the dependent relationship between secure key rate and sending length under three cases where security parameters are different and channel loss is $t = 15$ dB:

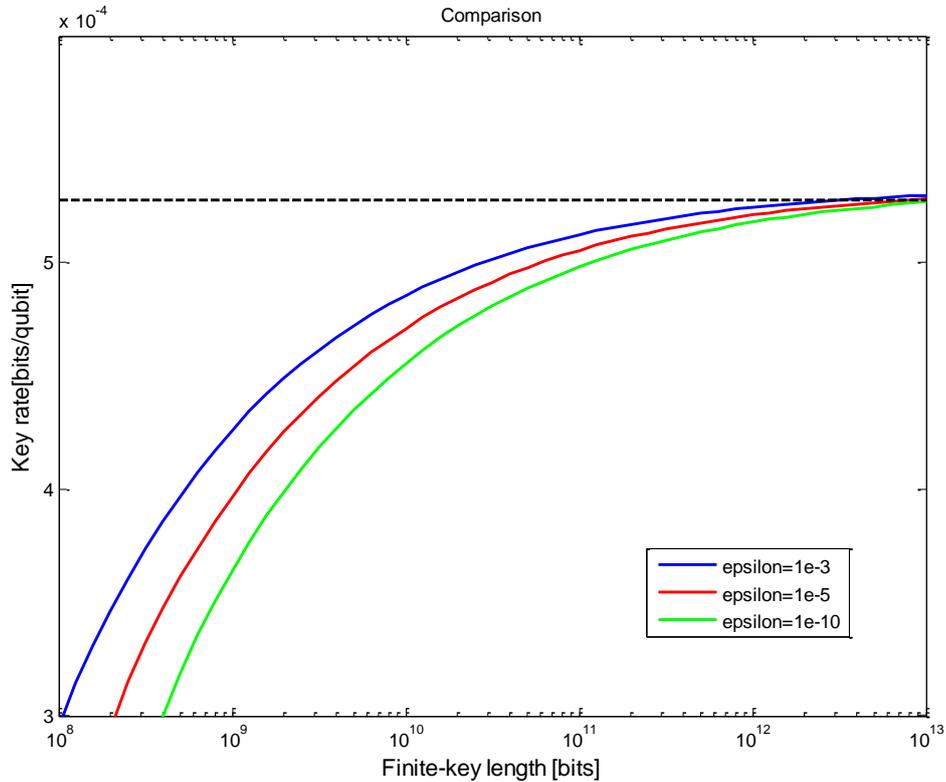

Figure 2: contrast diagram of secure key rate with different security parameters

In figure 2,the blue curve represents the secure key rate under the conditions that security parameter of final key is $10^{-3}$ with finite-key length, the red one represents $10^{-5}$, and the green one represents $10^{-10}$, while the black spotted curve represents the key rate with infinite-key length. It shows that when the finite-key length is $10^{13}$, the ratios between key rate deviation resulted from related security parameters and the one with infinite-key length are respectively 0.06%、0.14% and 0.23%, which is consist with the result confirmed in figure 1. So we demonstrate that a right length exits which realizes that the influence of security parameter on secure key rate can be ignored.

**4 Fair efficiency comparisons**

4.1 Importance

It is the aim of fair efficiency comparison to analyze the main element influencing secure key rate on the finite-key condition deeply through comparison, and the significance is to provide reference for promoting further secure key rate of decoy-state QKD protocols with biased basis choice. In particular, for the same sending length and UC security parameter of final key, our work provides a reference platform of fair efficiency comparison of decoy-state QKD protocols under UC security standard. Besides, through the fair efficiency comparison of typical protocols, we verify the main elements influencing secure key rate and search for optimization methods to improve it.

4.2 Conditions

When we give the fair efficiency comparisons of typical decoy-state QKD protocols, we need to analyze the following conditions besides the same sending (or receiving) length, UC security parameter of final key and statistical fluctuation method.

（1）secure key rate formula: mainly include the one based on GLLP formula and the one based on smooth entropy, and we will analyze the influence on secure key generation rate between different formulas in the fair efficiency comparisons;

（2）light intensity choice: the way of generating key in the Z (X) basis of signal state, in both Z and X basis of signal state , in Z (X) basis of all pulses and more, and their influence will be analyzed in the comparisons, too.

（3）the single bit error rate estimation method: we tend to estimate by signal state or decoy state, and similarly their influence will be analyzed in the comparisons.

（4）population sample of parameter estimation choice: include the number of sending light pulses (see in Ref.[17,24]) and receiving light pulses (see in Ref.

[22,32,33]) , and they are supposed to be the same when giving comparisons.

Other parameters remain unchanged apart from the above conditions. In addition, we don't limit the consistency in the comparison procedure towards parameters adjusting changes in the numerical optimization course, such as the ratio of basis bias choice, the radio of light intensity and security parameters adjusting in each postprocessing stage, so we note that the fair comparison in this article is relatively fair.

4.3 Comparisons towards schemes based on different formula of secure key rate

So far, we haven't seen fair efficiency comparison towards schemes based on GLLP formula and smooth entropy. Here taking decoy-state BB84 protocol with biased basis choice in the Z basis of signal state, we give a fair efficiency comparison between protocols based on GLLP formula and the ones based on Devetak-Winter bound and smooth entropy under UC security standard for the first time.

The fair comparison preconditions are assumed as: UC security parameters of final key are $\varsigma_{BB84} = 10^{-10}$ in them. Meanwhile, we choose the signal state's light intensity larger than decoy state, and estimate the single bit error rate by decoy state, where decoy-state modes are weak and vacuum decoy state. Statistical fluctuation method is standard error analysis, and population sample of parameter estimation choice is the number of sending light pulses. For the sake of convenience, we just consider dark count rate, $1.7 \times 10^{-6}$, detector efficiency, $4.5\%$ ,detector error rate, $3.3\%$ ,and channel loss , $t \in [0,30]$, （单位……）in practical systems.

Change $q_z^2$ into $q_z$ in （3）, and （2） and （3） are respectively formula of secure key rate in two schemes. After numerical optimization we can get figure 3 showing the contrast of two schemes when finite-key length $N \in \{6 \times 10^9, 1 \times 10^{12}, 1 \times 10^{15}\}$. Here GLLP represents the scheme based on GLLP formula, and SE represents the one based on smooth entropy.

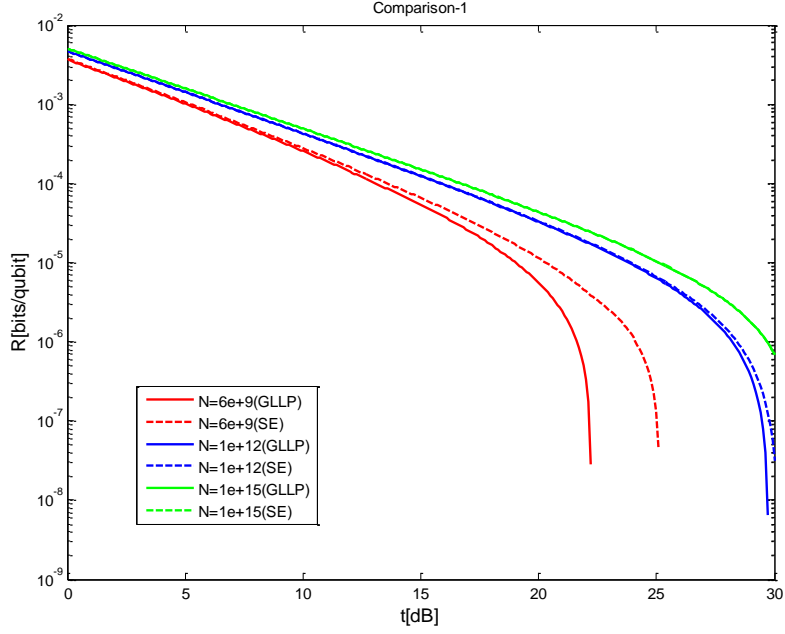

Figure 3 contrast diagram of secure key rate in two schemes with biased basis choice

After numerical optimization, when the finite-key length is $6\times10^9$, secure key rate of the protocol based on Devetak-Winter bound and smooth entropy is higher than the one based on GLLP formula. The degree of rate increase with three key length is shown in table 2.

Table 2 ratios of rate increase

| channel loss\finite-key length | 5 | 10 | 15 | 20 |
|---|---|---|---|---|
| $6\times10^9$ | 4.09% | 8.82% | 22.12% | 104.30% |
| $1\times10^{12}$ | 0.24% | 0.45% | 0.86% | 1.79% |
| $1\times10^{15}$ | 0.01% | 0.01% | 0.02% | 0.04% |

We can see in table 2 that when the length is longer, the degree of rate increase is smaller and two curves of secure key rate in two schemes is closer, which almost coincide when $N=1\times10^{15}$. It fully verifies that two schemes are almost equivalent under the infinite-key condition.

The main reason for higher secure key rate of protocol based on smooth entropy than the one based on GLLP formula with finite-key length is that the former needn't meet the precondition de Finetti's theorem requires, so the key rate formula is tighter. Another reason is resulted from security parameter. The UC security parameter of final key is given by $\varsigma_{BB84}=10^{-10}$, while the security parameter demanded in the postprocessing stage of protocol based on GLLP formula is higher, so its secure key generation rate is lower according to the conclusion in chapter 3 that when the security parameter is smaller, the deviation of statistical fluctuation and secret key bits consumed in each postprocessing stage is larger. Another advantage of the protocol based on smooth entropy is that we can give the secure key length which can be distilled in the privacy amplification stage through small smooth entropy viewed from Information Theory. Therefore, we can remove the secure key information acquired by Eve in a better way.

4.4 Comparisons towards schemes based on different light intensity choices

On the base of research in Ref. [22], we compare the secure key rate in schemes generating key in all light intensity（UC-Both）and the ones in single signal state.

Raymond protocol chooses the signal state's light intensity smaller than decoy-state, and sets the probability of Alice choosing X basis is 1 when sending decoy state, making the decoy-state detection rate higher and the degree of influence of statistical fluctuation to carry on parameter estimation by decoy state lower. However, according to Ref. [17], only when decoy-state light intensity is smaller is estimation accuracy better, so there exists a negative effect. Besides, UC security parameter of final key in Raymond protocol is $\varsigma_{Ray} = 10^{-5}$, isn't very secure.

The preconditions of fair comparison are set as: the UC security parameters of final key are $\varsigma_{UC} = 10^{-10}$, and we choose the signal state's light intensity larger than decoy-state. Bias basis setting is that the probability of choosing Z basis is larger than that of X basis but not equal to 1, whichever light intensity is sent. And statistical fluctuation method is Law of Large Numbers, and population sample of parameter estimation is the number of sending light pulses. Besides we still use ways in Raymond protocol in single bit error rate estimation method, decoy-state modes and parameter value choices in practical systems.

The formula of secure key rate in UC-Both protocol is:

$$R_{Both} \geq P_\mu Q(\mu) q_z^2 [S_{\xi_\mu}(A|E,\mu) - \Delta(n_\mu) - \frac{leak_{EC}(e_Z(\mu))}{n_\mu}]$$
$$+ P_v Q(v) q_z^2 [S_{\xi_v}(A|E,v) - \Delta(n_v) - \frac{leak_{EC}(e_Z(v))}{n_v}]$$
（5）

where the designing process, define of related parameter and statistical fluctuation method with finite-key length is referred in Ref. [37].

The finite-key length $N \in \{1\times 10^{12}, 1\times 10^{15}\}$, and through numerical optimization we can get the contrast diagram of secure key rate as figure 4.

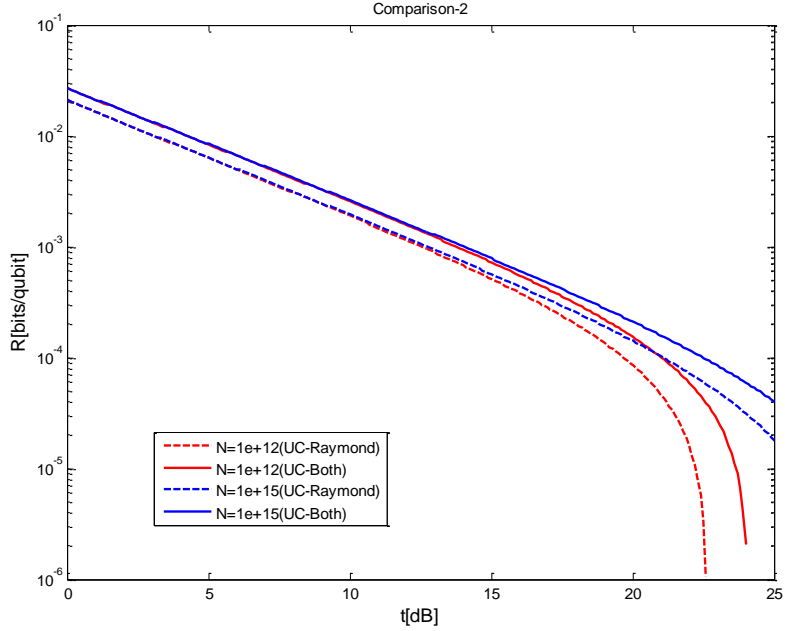

Figure 4 fair efficiency comparisons towards related schemes based on different light intensity choices

Through fair efficiency comparison we can come to the conclusion that though the key length is comparatively long ($N = 1 \times 10^{15}$), the secure key rate of UC-Both protocol, rising 35% on the base of the original, is always better than UC-Raymond protocol, which shows clearly that involving decoy state in key generation can increase secure key generation rate to a degree.

4.5 Comparisons towards schemes based on different single bit error rate estimation methods

Taking T12 protocol in Ref. [32] for an example, we verify the influence on secure key rate of the protocols between different single bit error rate estimation methods. Toshiba claimed that T12 protocol is the experimental implementation of the highest key generation rate currently. The original protocol generates key in both Z and X basis, raising secure key rate largely, meanwhile, it uses X basis part of signal state to carry on the single bit error rate estimation, and its advantage is decreasing the influence of statistical fluctuation. However, you must open the measuring information of X basis in the bit error rate estimation, but T12 protocol don't make it plain that how to solve the problem resulted from that X basis is not only involved in generating key but also sample estimation.

On the base of T12 protocol, we define that conducting the single bit error rate estimation by signal state is TS method, and by decoy state is TD method. The preconditions of this kind of fair comparison is set as: the UC security parameters of final key are $\varsigma_{Toshiba} = 1 \times 10^{-10}$, and decoy-state modes are two weak decoy states, satisfying $v_1 \geq v_2 \geq 0$. And population sample of parameter estimation choice is the number of receiving light pulse, and statistical fluctuation method is standard error analysis. Besides, we still use ways in T12 protocol in key generation ways, light intensity choice, bias basis setting, and parameter choice in practical systems.

The finite-key length $N \in \{1\times 10^{12}, 1\times 10^{15}\}$, and through numerical optimization we can get the contrast diagram of secure key rate between two estimation methods as figure 5(we use figure 5.1, 5.2, 5.3 and 5.4 to denote the above left, the above right, the lower left and the lower right subgraphs), where the horizontal axis is transmission distance, and the vertical axis is secure key rate.

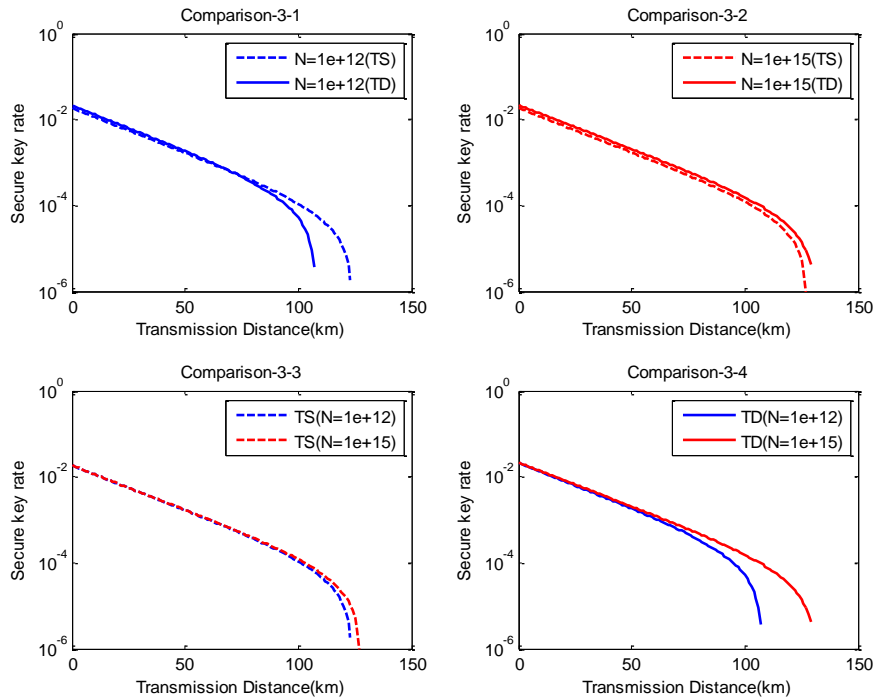

Figure 5 fair efficiency comparisons of related schemes with different estimation methods

We can see from figure 5.1 that when $N = 1\times 10^{12}$, these two different estimation methods have their own advantages respectively in raising key rate under the condition of different transmission distances. Taking figure 5.1 as an example, specific quantitative analysis is shown in table 3.

Table 3 efficiency comparison in T12 protocol based on different single

bit error rate estimation methods

| Transmission distances（km）\key rate （Mbps） | signal state's estimation (the original protocol) | the second decoy-state estimation | estimation involved in both signal state and decoy state |
|---|---|---|---|
| 35 | 3.506 | 3.941 | 3.941 |
| 50 | 1.710 | 1.869 | 1.869 |
| 65 | 0.823 | 0.850 | 0.850 |
| 80 | 0.382 | 0.351 | 0.382 |

Table 3 shows the the condition of key rate value choice got between two different estimation methods referring to four typical distance points chosen in T12 protocol, and the last column is the results of comprehensive consideration. In combination with figure 5.1, we can see that along with the distance increase causing the increase of channel loss, the parameter estimation sample decreases, and so the degree of the influence of statistical fluctuation rises. Under the condition of estimating single bit error rate based on the second decoy state and small channel loss, we get the higher key rate than TS method, and this fact demonstrates that the effect using decoy state to estimate is better. However, when channel loss is larger (e.g. 80km-120km), the key rate in TD protocol is lower than TS, and the reason is that at this moment the sample amount of signal state in estimation is larger than the second decoy state, and so the degree of the influence of statistical fluctuation is small, making the upper bound of single bit error estimation tighter and key rate higher than the latter. We can conclude from the last column that the method of choosing the tightest estimation value because of the comprehensive consideration of signal state and decoy state involved in estimation is better than the first two.

Figure 5.2 shows that when $N = 1 \times 10^{15}$, the effect of using TD method is always better than TS method, which shows when the sending length is long, we will get more accurate result by using decoy state to estimate.

Figure 5.3 and 5.4 reveal contrast diagrams of secure key rate when choosing TS and TD methods respectively with different key lengths. We can find that when we choose signal state to estimate, the secure key rate of two lengths is always close. However, when we choose decoy state, along with channel loss increasing, the difference of two lengths' secure key rate is larger and larger. As a result, we verify further that in resistance to the statistical fluctuation, the method of using signal state to estimate is better than decoy state.

After the analysis of four figures in figure 5, in the decoy-state QKD protocols with biased basis choice and under the condition of different sending lengths and channel losses, the influence on secure key rate of single bit error rate estimation methods has their own advantages, and therefore, we can promote the secure key rate further by means of comprehensive consideration in practical QKD systems.

Through above three aspects' analysis of fair efficiency comparison, we provide a standard and platform of efficiency comparison for typical decoy-state protocols with finite-key length, and on the base of fair efficiency comparison, we analyze and verify the major elements influencing secure key rate. At the same time we find the optimization methods to promote secure key rate of the protocol.

## 6. Conclusion

Towards the problem that security standards are different in decoy-state QKD protocols, we research the unified quantification of security parameter between decoy-state protocols based on GLLP formula and the ones based on Devetak-Winter and smooth entropy under UC security standard, and for the first time we give a fair efficiency comparison between two kinds of protocols. Besides, the impact of different sending length and secure parameters on secure key rate is shown, and we investigate the ways for improving secure key rate. The fair efficiency comparison mainly aims at typical protocols based on different formula of secure key rate, key generating part and the single bit error rate estimation methods. When meeting the prerequisite of fair comparison, through numeral optimization, the secure key rate of protocols based on Devetak-Winter bound and smooth entropy is better than the ones based on GLLP formula, and the method generating key in all light intensity can improve secure key generation rate further. Different single bit error estimation methods have advantages and disadvantages, and through comprehensive consideration we can improve secure key rate further.

**Acknowledgement**

This work is supported by the National High Technology Research and Development Program of China Grant No.2011AA010803, the National Natural Science Foundation of China Grants No.61472446 and No.U1204602 and the Open Project Program of the State Key Laboratory of Mathematical Engineering and Advanced Computing Grant No.2013A14.

参考文献

[1] Mayers D 1996 *CRYPTO'96, LNCS* 1109 343-357

[2] Mayers D 2001 *J ACM* 48 351

[3] Mayers D, Yao A 1998 *Proceeding FOCS '98 Proceedings of the 39th Annual Symposium on Foundations of Computer Science* (IEEE Computer Society Washington DC) pp503

[4] Bennett C H, DiVincenzo D P, Smolin J A, Wootters W K 1996 *Phys. Rev. A.* 54 3824

[5] Lo H K, Chau H F 1999 *Science* 283 2050

[6] Shor P, Preskill J 2000 *Phys. Rev. Lett.* 85 441

[7] R. Renner, Ph.D. thesis, Swiss Federal Institute of Technology (2005)

[8] Gottesman D, Lo H K, Lütkenhaus N, Preskill J 2004 Q. I. C. 4 5 325-360

[9] Bennett C H, Brassard G 1984 *In Proceedings of the IEEE International Conference on Computers, Systems and Signal Processing* (New York: IEEE Press) pp175–179

[10] M. Ben-Or, M. Horodecki, D. W. Leung, D. Mayers, and J. Oppenheim, in Second Theory of Cryptography Conference TCC2005, LectureNotes inComputer Science (Springer-Verlag, 2005), Vol. 3378, p. 386

[11] R. Renner and R. König, in Second Theory of Cryptography Conference TCC 2005, Lecture Notes in Computer Science (Springer-Verlag, 2005), Vol. 3378, p. 407

[12] R. Canetti, Tech. Rep. TR01-016, Electronic Colloquium on Computational Complexity (ECCC) (2001), preliminary version in IEEE Symposium on Foundations of Computer Science, pp. 136, 2001

[13] R. Canetti and H. Krawczyk, in EUROCRYPT 2002: Lecture Notes in Computer Science (Springer-Verlag, New York, 2002) Vol. 2332, p. 337 .

[14] Hwang W Y 2003 *Phys. Rev. Lett.* 91 057901

[15] Wang X B. Beating the photon-number-splitting attack in practical quantum cryptography. *Phys. Rev. Lett.* 2005, 94 : 230503

[16] Lo H K, Ma X F, Chen K 2005 Phys. Rev. Lett. 94 230504

[17] Ma X F, Qi B, Zhao Y, Lo H K 2005 Phys. Rev. A. 72 012326

[18] Wang X B, Peng C Z, Pan J W 2007 Appl. Phys. *Lett.* 90 3 031110

[19] Scarani V, Renner R 2008 *Phys. Rev. Lett* 100 20 200501

[20] Quan D X, Pei C X, Zhu C H, Liu D 2008 Acta Phys. Sin. 57 5600-5604 (in Chinese) [权东晓，裴昌幸，朱畅华，刘丹 2008 物理学报 57 5600-5604]

[21] Wang X B, Yang L, Peng C Z 2009 *New J. Phys.* 11 075006

[22] Cai R Y Q, Scarani V 2009 *New J. Phys.* 11 4 045024

[23] Zhou Y Y, Zhou X J 2011 Acta Phys. Sin. 60 100301 (in Chinese) [周媛媛，周学军 2011 物理学报 60 100301]

[24] Wei Z C, Wang W L, Gao M, Ma Z, Ma X F 2013 *Sci Rep.* 3 2453

[25] Rolando D S, Richard J H 2013 *Phys. Rev. A.* 87 062330

[26] Lucamarini M, Patel K A 2013 *Opt. Express* 10 1364

[27] Lucamarini M, Patel K A, Dynes J F, Fröhlich B, Sharpe A W, Dixon A R, Yuan Z

L, Penty R V, Shields A J 2013 Opt. Express 21 24550

[28] Fröhlich B, Dynes J F, Lucamarini M, Sharpe A W, Yuan Z L, Shields A J 2013 Nature 501 69

[29] Mafu M, Garapo K, Petruccione F 2013 Phys. Rev. A. 88 062306

[30] Mertz M, Kampermann H, Shadman Z, Bruß D 2013 Phys. Rev. A. 87 012315

[31] Krapick S, Stefszky M S, Jachura M, Brecht B, Avenhaus M, Silberhorn C 2014 Phys. Rev. A. 89 012329

[32] Lucamarini M, Patel K A, Dynes J F, Fröhlich B, Sharpe A W, Dixon A R, Yuan Z L, Penty R V, Shields A J 2013 Opt. Express 21 21 24550-24565

[33] Wen Lim C C, Curty M, Walenta N, Xu F H, Zbinden H 2014 Phys. Rev. A. 89 022307

[34] Fung C F, Ma X F, Chau H F 2010 Phys. Rev. A. 81 012318

[35] I. Devetak, A. Winter, Proc. R. Soc. A 461, 207 (2005)

[36] Cover M, Thomas J A 1991 Elements of Information Theory (New York Wiley Series in Telecommunications) pp106-120

[37] Li H X, Gao M, Ma Z, Ma C G, Wang W. Research on UC secure finite-key protocol with decoy state. Journal of Cryptologic Research, 2014, 1(6): 589–601.